\documentstyle[12pt,cite,psfig]{article}
\begin{document}
\begin{center}{\Large  \bf
Influence of Static and Dynamic Disorder \\
on the Anisotropy of Emission\\
in the Ring Antenna Subunits \\
of Purple Bacteria Photosynthetic Systems}
\\
\vspace{.5cm}
{\bf Pavel He\v{r}man}\\
{ \it
Department of Physics,  University of Hradec Kr\'{a}lov\'{e},
V. Nejedl\'{e}ho 573,\\
CZ-50003 Hradec Kr\'{a}lov\'{e}, Czech Republic\\
e-mail: pavel.herman@uhk.cz }

\vspace{.5cm}

{\bf Ulrich Kleinekath\"{o}fer }\\
{ \it Institut f\"{u}r Physik,  Technische Universit\"{a}t, D-09107
  Chemnitz, Germany \\
e-mail: kleinekathoefer@physik.tu-chemnitz.de}

\vspace{.5cm}
{\bf Ivan Barv\'{\i}k} \\
{ \it
Institute of Physics, Charles University,
Ke Karlovu 5,\\
CZ-12116 Prague, Czech Republic\\
e-mail: barvik@karlov.mff.cuni.cz}

\vspace{.5cm}

{\bf Michael Schreiber }\\
{ \it Institut f\"{u}r Physik,  Technische Universit\"{a}t, D-09107 Chemnitz,
  Germany\\
e-mail: schreiber@physik.tu-chemnitz.de}

\end{center}
\vspace{2.5cm}
\newpage
\begin{abstract}
  \baselineskip 3.6ex Using the reduced density matrix formalism the time
  dependence of the exciton scattering in light-harvesting ring systems of
  purple bacteria is calculated. In contrast to the work of Kumble and
  Hochstrasser (J.  Chem.\ Phys.\ 109 (1998) 855) static disorder
  (fluctuations of the site energies) as well as dynamic disorder
  (dissipation) is taken into account.  For the description of dissipation
  we use Redfield theory in exciton eigenstates without secular
  approximation. This is shown to be equivalent to the Markovian limit of
  \v{C}\'{a}pek's theory in local states.  Circular aggregates with 18
  pigments are studied to model the B850 ring of bacteriochlorophyls within
  LH2 complexes.  It can be demonstrated that the dissipation is important
  for the time-dependent anisotropy of the fluorescence.  Smaller values of
  static disorder are sufficient to produce the same decay rates in the
  anisotropy in comparison with the results by Kumble and Hochstrasser.
\end{abstract}

\newcommand{\be}{\begin{equation}}
\newcommand{\ee}{\end{equation}}
\newcommand{\bea}{\begin{eqnarray}}
\newcommand{\eea}{\end{eqnarray}}
\newcommand{\ba}{\begin{array}}
\newcommand{\ea}{\end{array}}
\newcommand{\ok}{\omega_k}
\newcommand{\skp}{\sin{\frac{k}2}}
\newcommand{\svisle}{|G_k^1-G_k^2|^2}
\newcommand{\bose}{[1+2n_B(\hbar \ok)]}
\newcommand{\podil}{\frac{\cosh(w)}{\sinh(w)}}
\newcommand{\odmocnina}{\sqrt{1-u^2}}
\newcommand{\imag}{ {\rm i } }
\newcommand{\ddt}{\frac{\text d}{{\rm d} t}}

\newpage

\section{Introduction}
The physical motivation of the present paper is a growing interest in the
exciton transfer in the antenna systems (ASs) of the purple bacteria
photosynthetic units (PSUs).  The light-harvesting complexes that are
present in photosynthetic systems perform two major functions: harvest
(absorb) the incident light and transport the energy in form of Frenkel
excitons to the reaction center (RC).  Many investigations, both
experimental and theoretical ones, have been directed towards understanding
of the exciton transfer in the ASs of purple bacteria PSUs and are reviewed
in \cite[and references therein]{pull,jpch,kuehn}.  Although considerable
progress has been made during recent years, our knowledge about the
mechanism of energy transfer and relaxation is still far from complete.

In 1995 the first high-resolution three-dimensional X-ray structure of a
bacterial antenna complex LH2 from {\it Rhodopseudomonas  acidophila}
was published by Mc Dermott et al.  \cite{cog1,cog2}.  It consists (see
Fig. 2 in \cite{cog2}) of nine identical units, protomers, each of which is
formed by two proteins ($\alpha$ and $\beta$) with bound
bacteriochlorophyls (BChl).  The nine $\alpha$ subunits are packed in an
inner ring to form a hollow cylinder of radius 1.8 nm.  The 9
$\beta$ subunits are arranged radially outwards with respect to the
$\alpha$ subunits to form another ring with a radius of 3.4 nm.  The
protein serves as a scaffold for the BChls and furthermore specifically
influences the spectroscopic properties of the BChls by supplying a
characteristic environment for them.  A ring with a radius of about 2.5 nm
of 18 BChl molecules is sandwiched between the $\alpha$  and
$\beta$ subunits.

The very symmetric arrangement with short distances between the pigments
provides a new impulse to the discussion about, e.g., the exciton transfer
regime and the exciton delocalization in LH2.  The extent of the exciton
delocalization, which could be reduced by dynamic and static disorders, has
been discussed. The principal question reads: Are the states contributing
to the optical and transport properties of the LH2 ring localized or
delocalized?

In the past knowledge of the energy transfer was mainly derived from steady
state spectroscopic experiments leading to several absorption bands at
different wavelengths.  Pigment molecules B850, that we are dealing with,
are characterized by the absorption wavelength in the LH2 subunits of the
ASs of purple bacteria.  At room temperature the solvent and the protein
environment fluctuate with characteristic time scales ranging from
femtoseconds to nanoseconds.  The dynamical aspects of the system are
reflected in the line shapes of the electronic transitions.  To fully
characterize the line shape of a transition and thereby the dynamics of the
system, one needs to know not only the fluctuation amplitude $ \Delta$
(coupling strength) but also the time scale of each process involved.  The
observed linewidths reflect the combined influence of static disorder and
exciton coupling to intermolecular, intramolecular, and solvent nuclear
motions.  The simplest approach is to decompose the line profile into
homogeneous and inhomogeneous contributions. In more sophisticated models,
each process is defined with its characteristic time scale as well as a
coupling strength \cite{barmit,barchemphys,knoester1,knoester2}.

Both the exciton-phonon coupling and the static disorder lead to
localization of excitons. One tries to avoid the influence either of the
dynamical disorder by measuring at very low temperatures or of the static
disorder by performing single-molecule spectroscopy experiments.  The
absorption, circular dichroism, and hole-burning spectra of the LH2 complex
are reproduced reasonably well by theoretical treatments, that consider
Frenkel exciton states of the entire complex but invoke moderate disorder
in the excitation energies of the individual BChls \cite{pull}.

In a homogeneous system, in which the $Q_{y}$ transition dipole moments of
the B850 BChls all lie approximately in the plane of the ring, most of the
dipole strength of the B850 band would come from a degenerate pair of
orthogonally polarized transitions at an energy slightly higher than the
transition energy of the lowest exciton state.  However, energetic or
structural disorder that disrupts the symmetry of the complex will make
both the low-energy transition and transitions at higher energies weakly
allowed.

Time-dependent experiments (absorption, fluorescence, photon echo, etc.),
which used  (sub-)picosecond light pulses of low energy with high
repetition rate tunable through the infrared absorption bands of the
various BChl pigments in a variety of bacterial ASs, made it possible to
study the long-time as well as femtosecond dynamics of the energy transfer
and relaxation \cite{pull,freib1,freib2,freib3,freib4,koolhaas}. The nature
of the rapid relaxation is of interest because it depends on how
photosynthetic antenna complexes absorb and transfer energy and because it
may provide an experimental window to protein dynamics on very short time
scales \cite{nagarjan3}.

The interpretation of time-dependent experiments on femtosecond time scale
requires a theory which incorporates both dynamic disorder and different
kinds of static disorder.  Recently, models based on extended exciton
states with moderate static disorder $\Delta \approx J/2$ have found
acceptance in the interpretation of the experiments
\cite{freib1,freib2,freib3,freib4,koolhaas, flem1,muk,grond2,nov5,nov6} in
which it has been shown that the elementary dynamics occurs on a time scale
of about 100 fs \cite{nagarjan1,nagarjan2,nagarjan3,nagarjan4}.  For the
ring of BChls in the LH2 antenna complex Kumble and Hochstrasser
\cite{hoch} has presented a time-domain analysis of the effects of local
energy inhomogeneity upon the dynamics of optical excitations. They
examined the manifestation of disorder scattering in polarized femtosecond
spectroscopy and the degree to which exciton delocalization is revealed in
emission and transient absorption anisotropy measurements.  The time
evolution of the quantum states, prepared by impulsive excitation of a
statically disordered circular aggregate model for LH2 antenna complexes,
have been calculated exactly quantum mechanically for varying degrees of
inhomogeneity (static disorder), but without taking into account the
interaction with a heat bath (dynamic disorder).  For a Gaussian
distribution of local energies, the dynamics of coherence loss (scattering)
has been  explored as a function of the ratio of the standard deviation $\Delta$
of the distribution to the intersite interaction energy $J$.  (Please note
that Kumble and Hochstrasser used the symbol $\sigma$ for the standard
deviation in their paper \cite{hoch}.) It has been found, that modest
degrees of disorder ($\Delta/J \approx 0.4-0.8$) are sufficient to cause
scattering on a sub-100 fs time scale.  The estimate of static disorder
emerging from this investigation should be considered as an upper limit. By
neglecting the influence of phonons, Kumble and Hochstrasser's model does
not take into account the dynamic scattering effects which contribute to
dephasing of the initial wave packet state and which promote thermalization
of the dephased populations.

Nagarjan et al.\cite{nagarjan1} measured the changes in absorption and
stimulated emission caused by excitation with 35 fs pulses centered at 875
nm. Substantial relaxation on the time scale of 10-100 fs and an
anomalously large initial anisotropy of 0.7 in the bleaching and
stimulated-emission signal has been observed. These results have been
interpreted in the simple model of a homogeneous system in which
excitations are delocalized over the whole ring.  The high initial
anisotropy was ascribed to coherent excitation of a degenerate pair of
states with allowed transitions and the relaxations to electronic
dephasing and equilibration with states at lower energies which have
forbidden transitions. It has been shown in Ref.~\cite{nagarjan3} that
excitation of membrane-bound LH2 complexes with low-intensity femtosecond
pulses causes changes in absorption and stimulated emission that initially
depend on the excitation wavelength but relax to a quasi-equilibrium with a
time constant of 100~$\pm{} $~20~fs.  The signals have an apparent initial
anisotropy of approximately 0.5 when the complex is excited with broadband
pulses, and of 0.35-0.4 with narrower pulses. The anisotropy decays to 0.1
with a time constant of about 30~fs.

Using a density-matrix formalism Nagarjan et al.\cite{nagarjan3} have shown
that the initial light-induced signals are consistent with coherent
excitation of multiple exciton levels in an inhomogeneous ensemble of LH2
complexes and that the main features of the spectral relaxations and the
anisotropy can be explained by electronic dephasing and thermal
equilibration within the manifold of exciton levels.  For one initial state
three different possibilities of the form of the exciton density matrix
$\rho$ have been taken into account.  Beside the full matrix description,
which has been represented as a coherent superposition of the allowed "one
exciton" states, the second is obtained by setting the off-diagonal terms
of $\rho$ to zero, which gives an incoherent mixture of one-exciton states
with the same initial populations.  Finally, in the third form the diagonal
elements of $\rho$ are replaced by the populations for a Boltzmann equilibrium
at temperature T (with the off-diagonal terms zero).

Nagarjan et al.\cite{nagarjan3} have discussed several different processes
which could contribute to the spectral relaxations and the rapid decay of
the anisotropy. They concluded, that the main features of the spectral
relaxations and the decay of anisotropy are reproduced well by a model that
considers decay processes  of electronic coherences within the
manifold of the exciton states and the thermal equilibration among
excitonic states.  They have not attempted to calculate the
dynamics theoretically.  But the model described in \cite{nagarjan3}
appears capable of rationalizing the observation that the anisotropy decay
is faster than the spectral relaxation.

Moderate static disorder, which leads to electronic dephasing, causes a
large decrease in the calculated anisotropy, whereas the relaxation of the
overall signal occurs mainly during thermalization,  because the two
most strongly allowed exciton states tend to be close together in energy
and to have approximately orthogonal transition dipoles.  Nagarjan et al.
\cite{nagarjan3} concluded that the static inhomogeneity assumed in their LH2
model would cause the off-diagonal terms $\rho_{ab}$ and $\rho_{ba}$ of the
density matrix to decay with a time constant of the order of 160~fs. This
suggests that the more rapid dephasing indicated by the anisotropy decay
($\tau \approx 30$~fs) is driven mainly by dynamic processes.

The aim of the present paper is to extend the investigation by Kumble and
Hochstrasser \cite{hoch} and also by Nagarjan et al. \cite{nagarjan3} taking
into account the simultaneous influence of static and dynamic disorders on
the exciton scattering after impulse excitation and on the time-dependent
optical anisotropy.
Different time dependent spectroscopic
properties have been calculated before but within the secular approximation
\cite{kuehn97}.

The remainder of this article is organized as follows: In the next section
we present the model we have used. In section 3 several different theories
giving the dynamical equations for the exciton density matrix are reviewed.
Our results are presented in section 4 and in the last section we discuss
them and draw some conclusions.

\section{Model}

\subsection{Hamiltonian of the ideal rings}

In this paper we deal with just one exciton, already created at time $t=0$,
on a ring interacting with a heat bath.  The Hamiltonian therefore
consists of three parts:
 \begin{eqnarray}
   H^{0} &=&\sum_{m,n} J_{mn} a_m^{\dag} a_n +\sum_q\hbar\omega_q
   b_q^{\dag} b_q + \frac{1}{\sqrt{N}}\sum_m\sum_{q}G_{q}^m
   \hbar\omega_{q}a_{m}^{\dagger}a_{m} (b_{q}^{\dagger}+b_{-q})\nonumber \\
&=& H_{\rm ex}^{0}+H_{\rm ph}+H_{\rm ex-ph}. \label{he}
 \end{eqnarray}
 The first term
is the Hamiltonian $H_{ex}^{0}$ belonging to the single exciton in the
ideal ring.  Operators $a_m^{\dag}$ or $a_m$ create or annihilate an
exciton at site $m$.  $J_{mn}$ for $m\ne n$ is a transfer integral (also
called resonance integral) between sites $m$ and $n$.  The diagonal
elements $J_{nn}$ are the local energies $\epsilon_n$ at site $n$ which
could take into account, e.g., a dimerization $\epsilon_{2n} \neq
\epsilon_{2n+1}$.  The second term, $H_{ph}$, describes the phonons. We
assume independent heat baths for each chromophore and the harmonic
approximation for the phonons. $b_q^{\dag}$ and $b_q$ denote phonon
creation and annihilation operators, respectively.  The last term,
$H_{ex-ph}$, represents the interaction between the exciton and the bath.
The exciton--phonon interaction is assumed to be site--diagonal and linear
in the lattice displacements. The term $G^m_q$ denotes the
exciton--phonon coupling constant.

In the framework of the extended exciton states it is assumed that both
intra-dimer $J_{12}$ and inter-dimer $J_{23}$ transfer integrals are strong
enough to build exciton states spanning  the whole rings.  Inside one
ring the pure exciton Hamiltonian $H_{ex}^0$ (\ref{he}) can be diagonalized
using the wave vector  representation with corresponding
delocalized "Bloch" states and energies.  Considering, e.g., only nearest
neighbor transfer matrix elements $J_{mn}= -J
(\delta_{m,n+1}+\delta_{m,n-1})$, the same local energies $\epsilon_{n}$
at every site and using Fourier transformed excitonic operators (Bloch
representation)
\begin{equation}
a_k = \sum_n a_n \, e^{i k n} \quad , \quad k=\frac{2\pi}{N} \,
 l \quad,\quad l = 0,\pm{} 1 , \ldots \pm{} N/2
\end{equation}
the simplest exciton Hamiltonian  in $\vec k$ representation reads
\begin{equation}
\label{f1.7}
H_{ex}^0
= \sum_k E_k \, a_k^+ a_k~.
\end{equation}
The dispersion of the excitonic energies is given by
\begin{equation}
\label{f1.8}
E_k = - 2 \, J \, \cos k~.
\end{equation}
A splitting of the local degenerate exciton energies
in one ring, to a band of  energies $E_k$, corresponding
to the exciton eigenstates,
is produced.

\subsection{Hamiltonian of the static disorder}

The quasistatic disorder is described by
the time-dependent part of the Hamiltonian $H = H^{0} + H_{1}(t)$. This part
\begin{equation}
H_{1}(t) = \sum_{n} \epsilon_{n}(t) a_{n}^{+}a_{n}
\end{equation}
\noindent
describes fluctuations of the local excitation energies $\epsilon_{n}(t)$
due to slow motion of the protein environment.
Fluctuations of the local excitation energies $\epsilon_{n}(t)$ have
exponentially decaying correlation functions.
The mean values and correlation functions are given by
$$
\langle \epsilon_{n}(t) \rangle  = 0,
$$
\begin{equation}\langle \epsilon_{n}(t)\epsilon_{m}(\tau) \rangle
  = \delta_{mn} \Delta^2 \exp(-\lambda (t-\tau)).
\end{equation}
In what follows the simplification is used, that we work only with pure
static disorder, $\lim \lambda \to 0 $, and a Gaussian distribution for the
uncorrelated local energy fluctuations $ \epsilon_{n}^{s}$ with a standard
deviation $\Delta$.

\subsection{Microscopic parameters }

Despite the long history of theoretical and computer modeling of the
experimental results no unified set of microscopic parameters entering the
Hamiltonian $H$ has been revealed up to now
\cite{schulten97,freib1,freib2,freib3,freib4,koolhaas,flem1,muk,grond2,nov5,nov6}.

The one-exciton band of the ideal ring $E(k)$ consists of two groups of 9
transitions arranged asymmetrically relative to the zero local site energy.
These Davydov manifolds appear because the 18 BChl molecules in the B850
ring are grouped into heterodimers with different coupling energies within
the elementary dimers $J_{12}$ and between them $J_{23}$ as well as
different local energies $ \epsilon_{1} \neq \epsilon_{2}$.

In the presence of static disorder ($\Delta \neq 0$ in $H_{1}$) virtually
all exciton transitions gain strength at the expense of the low energy $ k
= \pm{} 1 $ transitions, which dominate the spectrum of the ideal B850 ring.
There is still a discrepancy between the measurement by Freiberg et
al.~\cite{freib2} and results of hole-burning experiments by Small (see
references in \cite{pull}).  In the former experiments considerable
intensity of about $13 \%$ (averaged over a large ensemble) attained by the
lowest $k = 0$ transition has been revealed. In interpreting these
experiments it was concluded \cite{freib2} that the shape of the linear
absorption spectrum of B850 at 8~K could be well reproduced by taking
$\Delta \approx 0.7$. The hole burning results at 4~K imply much smaller
disorder leading to a relative intensity of the $k = 0$ subband of only $ 3
\% $.  The model parameters obtained by Freiberg from fitting  the
simulated linear and nonlinear absorption spectra to the low temperature
experimental data are the transfer integrals $J_{12} =
375$~cm$^{-1}$ and $J_{13} =20-30$~cm$^{-1}$ as well as the strength of the
static disorder $\Delta = 216$~cm$^{-1}$ with ${\rm FWHM} =507$~cm$^{-1}$.
The dynamic disorder leads to a broadening of
the exciton  levels $E(k)$ with a ${\rm FWHM} = 58$~cm$^{-1} =
J/6.5$.

Novoderezhkin et al.\ \cite{nov5,nov6}  fitted the
pump-probe spectra of the LH2 antenna. Taking into account the difference
between the local BChl energies $ \Delta \epsilon = \epsilon_1 - \epsilon _2 $
in the basic dimer which leads to the splitting of the exciton energy band
to two subbands, the following set of microscopic
parameters : the transfer integrals $J_{12} = 400$~cm$^{-1}$, $J_{23}=
290$~cm$^{-1}$, the
the homogeneous line widths of exciton states in the lower subband
$\Gamma_{1L} = 240$~cm$^{-1}$ and in the higher subband  $\Gamma_{1H} =
340$~cm$^{-1}$ and the strength of the static disorder $\Delta =
450$~cm$^{-1}$ has been used.
Koolhaas et al.\
\cite{koolhaas} have recently obtained the transfer integrals $J_{12} =
396$~cm$^{-1}$ and $J_{23}= 300$~cm$^{-1}$.

In what follows we will use the transfer integral $J_{12}$ as energy unit.
We choose $J_{23} = 0.7$, $\Delta \epsilon = \epsilon_1 - \epsilon_2 = J_{12}$,
and $\Delta$ between 0.2 and 1. To convert the time $\tau$ in these units
into seconds one has to divide $\tau$ by $2 \pi c J_{12}$ with $c$ the speed
of light in cm/s and $J_{12}$ in units of cm$^{-1}$.  So $\tau=1$
corresponds to 21.2~fs for $J_{12}= 250 ~{\rm cm}^{-1}$ and to 13.3~fs for
$J_{12}= 400 ~{\rm cm}^{-1}$.

\subsection{Nature of the initially prepared state }

For a coplanar arrangement of site transition moments $ \vec \mu{} _{n} $
dipole-allowed transitions from the ground state populate only the
degenerate $k= \pm{} 1$ levels of the ideal (without dynamic and static
disorders) ring with Hamiltonian $ H_{ex}^{0}$ and eigenstates $ |k\rangle
$.  If static disorder of the local energies is present ($ \Delta \neq 0
$), the stationary states, i.e.\ the eigenstates $| a\rangle$ of the
Hamiltonian $ H_{ex}^{0} + H_{1}$, correspond to mixtures of  $|k \rangle $
and an excitation
will prepare a superposition of the $|k \rangle$ states.

However, Kumble and Hochstrasser \cite{hoch} concluded, that in the case of
pump-pulse excitation the dipole strength is simply redistributed among the
exciton levels due to disorder. So the amplitudes of site excitations and
the phase relationships in the initial state are necessarily identical to
that of an equal superposition of  $ k=\pm{} 1$ excitons of the ideal ring.  Thus,
generally, the excitation with a pump pulse of sufficiently wide spectral
bandwidth will always prepare the same initial state, irrespective of the
actual eigenstates of the system. The nature of this initial state is
entirely determined by the selection rules of the unperturbed
system.

We shall use the following definitions for the dipole strength
 \begin{equation}
 \vec\mu{} _{a} = \sum_{n=1}^{N} c_{n}^{a} \vec \mu{} _{n}
\end{equation}
of state $|a\rangle$ of the
real system and for the dipole strength
 \begin{equation}
 \vec \mu{} _{\alpha} =\sum_{n=1}^{N} c_{n}^{\alpha} \vec \mu{} _{n}
 \end{equation}
of state $ |\alpha\rangle$ of
the ideal crystal.  The coefficients $c_{n}^{\alpha}$ and $c_{n}^{l}$ are
the expansion coefficients of the eigenstates of the ideal and disordered
aggregates in site representation.

The pump pulse is characterized by the polarization unit vector $ \vec
e_{i}$.
Then the initial condition for the density matrix by pulse excitation with
the polarization $\vec e_{i}$ is given by
(Eq. (1a) in \cite{nagarjan3}):
\begin{equation} \rho_{\alpha\beta} (t=0; \vec e_{i})   = 
\frac{1}{A}( \vec e_{i}\cdot{} \vec \mu{} _{\alpha})( \vec e_{i}\cdot{} \vec
\mu{} _{\beta}),
\end{equation}
where $A=\sum_\alpha ( \vec e_{i}\cdot{} \vec \mu{} _{\alpha})( \vec e_{i}\cdot{} \vec
\mu{} _{\alpha})$.
This initial condition for  pulse excitation \cite{hoch}
should be expressed in the eigenstate basis of the real system:
\begin{eqnarray} \rho_{ab} (t=0; \vec e_{i}) &=& \sum_{ \alpha} \sum_{ \beta}
\langle a| \alpha\rangle \rho_{\alpha\beta} (t=0; \vec e_{i})
\langle \beta | b\rangle\nonumber\\
& =&              
\frac{1}{A} \sum_{ \alpha} \sum_{ \beta}
( \vec e_{i}\cdot{} \vec \mu{} _{\alpha}) ( \vec e_{i}\cdot{} \vec \mu{} _{\beta})
\langle a| \alpha \rangle\langle  \beta | b\rangle,
\label{rhot}
\end{eqnarray}
>From (\ref{rhot}) one obtains
\begin{eqnarray} \rho_{ab} (t; \vec e_{i}) &=&
{\mathrm e}^{- \frac {i}{\hbar}  t (E_{a}-E_{b})} \rho_{ab} (t=0; \vec
e_{i})\nonumber\\
&=&
\frac{1}{A}{\mathrm e}^{-  \frac{i}{\hbar}  t (E_{a}-E_{b})}
\sum_{ \alpha} \sum_{ \beta}
( \vec e_{i}\cdot{} \vec \mu{} _{\alpha}) ( \vec e_{i}\cdot{} \vec \mu{} _{\beta})
\langle a| \alpha \rangle \langle \beta | b\rangle. \end{eqnarray}

\subsection{Static disorder-induced scattering of the initially pure state}

Pump-probe and spontaneous emission measurements on the femtosecond time
scale have revealed ultrafast depolarization in the core and peripheral ASs
with biphasic kinetics: a fast sub-100~fs component and a slower decay on a
200-400~fs time scale were observed in all cases \cite{pull}.  Kumble and
Hochstrasser \cite{hoch} thoroughly investigated the time dependence of the
disorder-induced scattering of the initial state $|k \pm{} 1 \rangle$.  The
evolution of the initially delocalized exciton
\begin{equation}
  P_{k}(t) = | \langle k | e^{-\frac{i}{\hbar}Ht} | k \rangle | ^2
\end{equation}
 has been followed considering only the influence of a static
uncorrelated distribution of the local site energies in a circular
aggregate.

In the  basis $ | k \rangle $ of the ideal ring the time evolution can be
described as scattering of the initially prepared  exciton $ |k =\pm{} 1
\rangle $ into those levels to which it is coupled by the presence of
disorder.  The nonexponential nature of scattering dynamics has been
demonstrated: for $\Delta = J/3.5$ the initial state decays in two distinct
phases; a dominant $ \approx 100$~fs component and a slower (200--300~fs)
decay are evident.

\subsection{Consequences of the static disorder-induced scattering
in optical anisotropy measurements}

Optical anisotropy measurements provide a convenient mean to follow
experimentally the decay of a virtual exciton within the LH2 ring.  Since
this initial state is prepared as a superposition of eigenstates of the
aggregate, polarized pump-probe (or fluorescence) signals contain
contributions from interference terms which last so long that
the phase coherence between different levels is maintained. The decay of
these cross terms directly reflects the time scale of electronic dephasing.
In the present case it therefore represents the survival time of a virtual
exciton within the disordered system.

Kumble and Hochstrasser \cite{hoch} calculated the usual time-dependent
anisotropy using
 \begin{equation}
 \label{aniso}
 r(t) = \frac { \langle S_{xx}(t)\rangle - \langle S_{xy}(t)\rangle} {
   \langle S_{xx}(t)\rangle + 2\langle S_{xy}(t)\rangle}
\end{equation}
where the brackets $ \langle \rangle$ denote the
ensemble average and the average over the direction of the laser pulses
with fixed relative directions $\vec e_1$ and $\vec e_2$.  For excitation
and probe pulses which are polarized along the $\vec e_x$ and $\vec e_y$
directions, respectively, the transient gain signal from the delocalized
initial state is given by \cite{hoch}
\begin{equation}
\langle S_{xy}(t) \rangle =
\langle | \sum_{\alpha,l,n} (\vec{e}_x\cdot{} \vec\mu{} _{\alpha})
(\vec{e}_y\cdot{} \vec\mu{} _{l}) c_{n}^{\alpha *}c_{n}^{l} {\mathrm e}^{-i \omega_l
  t}|^{2} \rangle
\end{equation}
where the indices $\alpha$ and $l$ label the
eigenstates of the virtual and disordered aggregates, respectively. The
index $n$ represents the individual site excitations.
The gain signal $\langle S_{xx}(t) \rangle$ is defined
accordingly.

In Kumble and Hochstrasser's modeling, the decay of the initial state has
been directly manifested itself as a depolarization reducing the anisotropy
from $0.7$ to $0.3-0.35$ and a
subsequent recurrence to establish a final value of $0.4$. Kumble and
Hochstrasser supposed that equilibration between eigenstates of the
aggregate would lead to a further decrease of the anisotropy to $\approx
0.1$, but this process has not been considered in their paper \cite{hoch}.
They estimated, by comparison of their results with  experiment,
that the static disorder in a LH2 ring has a value of $\Delta \approx
0.4-0.8$ which should be considered as an upper limit since dynamic
variations of the relative eigenstate energies occurring on a 100~fs time
scale would contribute to the scattering process.

\subsection{Effects of the dynamical disorder}

As pointed out in Sec.\ 1 the aim of the present paper is to extend the above
described theory given in Ref.~\cite{hoch} by including the effect
of dynamic scattering which contributes to dephasing of the initial wave
packet and promotes a thermalization of the dephased populations.  To deal
with the dynamic disorder, one  works within the exciton density
matrix formalism instead of using only the exciton wave functions.

Taken only those terms into account which are directly connected to the
polarization the time-resolved fluorescence $P_{\rm if}(\omega,t)$ with
initial $\vec e_{i}$ and final polarization $\vec e_{f}$ of the light is
given (cf.\ Chapter 3, Eq.  (2.58) in \cite{lin}) in our notation as
\begin{eqnarray} \label{trf}
   &P_{if}(\omega,t)=
 A\sum\limits_{l} \sum\limits_{ l'}
 \rho_{ll'}(t) ( \vec e_{f}\cdot{} \vec \mu{} _{l'}) ( \vec e_{f}\cdot{} \vec \mu{} _{l})
 [\delta(\omega-\omega_{l'0})+ \delta(\omega-\omega_{l0})] =& \nonumber\\ &
 A\sum\limits_{l} \sum\limits_{ l'}
\rho_{ll'}(t)
\sum\limits_{ k} \sum\limits_{ j}
( \vec e_{f}\cdot{} \vec \mu{} _{k}) ( \vec e_{f}\cdot{} \vec \mu{} _{j})
\langle j|l\rangle \langle l'|k\rangle [\delta(\omega-\omega_{l'0})+
\delta(\omega-\omega_{l0})]&  \label{pomega}
\end{eqnarray}
and in case of no interaction with a bath as
\begin{eqnarray}
   &P_{if}(\omega,t) = & \nonumber\\
& \sum_{ l} \sum_{ l'} \sum_{ \alpha} \sum_{ \beta}
{\mathrm e}^{- \frac {i} {\hbar} t (E_{l}-E_{l'})} ( \vec e_{i}\cdot{} \vec
\mu{} _{\alpha}) ( \vec e_{i}\cdot{} \vec \mu{} _{\beta}) \langle l| \alpha \rangle \langle
\beta | l'\rangle ( \vec e_{f} \cdot{} \vec \mu{} _{l'}) ( \vec e_{f}\cdot{} \vec \mu{} _{l})
\times{} & \nonumber\\
&[\delta(\omega-\omega_{l'0})+
\delta(\omega-\omega_{l0})].&
\end{eqnarray}
This expression can be directly rewritten (after integration
over $\omega$) in the form of Kumble and Hochstrasser:
$$
S_{ij}(t)=\int P_{if}(\omega,t)d\omega=
$$
$$
\int\sum_{ l} \sum_{ l'} \sum_{ \alpha} \sum_{ \beta}
{\mathrm e}^{- \frac {i} {\hbar} t (E_{l}-E_{l'})}
( \vec e_{i}\cdot{} \vec \mu{} _{\alpha}) ( \vec e_{i}\cdot{} \vec \mu{} _{\beta})
\langle l| \alpha \rangle \langle \beta | l'\rangle
( \vec e_{f} \cdot{} \vec \mu{} _{l'}) ( \vec e_{f}\cdot{} \vec \mu{} _{l}) $$
$$ \times{}
[\delta(\omega-\omega_{l'0})+
\delta(\omega-\omega_{l0})]d\omega=$$
$$\int\sum_{ l} \sum_{ l'} \sum_{ \alpha} \sum_{ \beta}
{\mathrm e}^{- \frac {i} {\hbar} t (E_{l}-E_{l'})}\times{} $$
$$( \vec e_{i}\cdot{} \vec \mu{} _{\alpha}) ( \vec e_{i}\cdot{} \vec \mu{} _{\beta})
\sum_{n}\sum_{n'}
\langle l|n\rangle\langle n| \alpha \rangle \langle \beta |n'\rangle\langle n'| l'\rangle
( \vec e_{f} \cdot{} \vec \mu{} _{l'}) ( \vec e_{f}\cdot{} \vec \mu{} _{l})
[\delta(\omega-\omega_{l'0})+
\delta(\omega-\omega_{l0})]d\omega =
$$
$$
 \left (\sum_{ \alpha} \sum_{ l} \sum_{ n}
( \vec e_{i}\cdot{} \vec \mu{} _{\alpha}) ( \vec e_{f}\cdot{} \vec \mu{} _{l})
\langle l|n\rangle\langle n|\alpha\rangle  {\mathrm e}^{-i \omega_{l} t}
\right )
\times{} $$
$$\left (\sum_{ \beta} \sum_{ l'}  \sum_{ n'}
( \vec e_{i}\cdot{} \vec \mu{} _{\beta}) ( \vec e_{f\cdot{} } \vec \mu{} _{l'})
\langle \beta|n'\rangle  \langle n'|l'\rangle {\mathrm e}^{i \omega_{l'} t}
\right )=$$
\begin{equation}
| \sum_{\alpha,l,n}  (\vec{e}_i\cdot{} \vec\mu{} _{\alpha})
                       (\vec{e}_f\cdot{} \vec\mu{} _{l})
            \langle\alpha|n \rangle \langle n|l \rangle  {\mathrm e}^{-i \omega_l
            t}|^{2}.
\end{equation}

We see, that the crucial quantity entering the time
dependence of the anisotropy is the exciton density matrix.
In the next section we shall therefore  review several methods leading to
dynamic equations for the exciton density matrix which include
microscopically the interaction of the exciton with the bath. They could be
obtained from, e.g., the general Tokuyama-Mori approach only by serious
approximations. One of them is the  Redfield model with and without the
so--called secular approximation \cite{redfield1,redfield2}, which has been
applied to exciton transfer and relaxation problems in AS rings several
times.  It has been shown \cite{cbh,egorova} that in contrast to the
justification of the  secular approximation, the
non-secular terms do not average out in time.

Our result, presented in  section 4,  have therefore been obtained
with the Redfield model without secular approximation.

\section{Dynamic equations for the reduced exciton density matrix}

The bath--related relaxation can be described in a variety of ways. Among
others these are the path integral methods \cite{weis99,makr98}, the
semigroup methods \cite{davi98b,kosl97,kohe97a}, and the reduced density
matrix (RDM) theory \cite{blum96}. Here we concentrate on the latter.

Having defined the Hamiltonian $H$ of the system, one exciton and a
bath, one has to solve the equations of motion for the complete
density matrix $\sigma$, namely the Liouville equation
\begin{equation}
i\hbar\frac{\partial}{\partial t}\sigma(t) =  L\sigma.
\end{equation}
This is not a simple task because one has to find the time development of
all matrix elements of the density matrix - diagonal and off-diagonal - in
any representation which takes into account the exciton and phonon states.

Nevertheless, this treatment is not necessary in many cases.
Information,
which is used in further investigations, is in many cases limited.  For
example, the site occupation probabilities $P_{m}(t)$ of the exciton
(the diagonal matrix elements in site representation after averaging over the
bath  variables) are in many cases the most interesting quantities
in theoretical investigations of the exciton transfer.  The transfer
problem of, e.g. a Frenkel exciton interacting with a  bath, is
usually treated by projecting out the bath degrees of freedom. This yields
dynamic equations for the single-exciton density matrix $\rho$
complementing the usual single-exciton Liouville equation by terms
describing the bath influence. The latter terms must, on the other hand, be
properly parametrized.

\subsection{Stochastic theories}

In the stochastic treatment of the exciton interaction with the dynamic and
static disorders the influence of disorder is described by a stochastic
process, i.e.\ the interaction part of the microscopic Hamiltonian is
replaced by a stochastic time-dependent model Hamiltonian.
In the original version \cite{reineker} of the stochastic Liouville
equation method (SLE) the bare exciton is influenced by a stochastic field
\begin{equation}
H_{2} (t) = \sum_{m,n} h_{mn} (t) \, a_{m}^{+} a_{n}.
\end{equation}

Grover and Silbey \cite{silbey} suggested that one could understand the
excitation in  molecular aggregates also as a fully dressed exciton (polaron)
obtained from the microscopic Hamiltonian using a canonical transformation.
The stochastic field acting on the dressed excitation has a different
microscopic meaning as that one acting on the bare exciton
\cite{capbar}.  The stochastic field obtained from the microscopic
Hamiltonian for the local and linear exciton-phonon interaction leads
\cite{capbar}, in the bare exciton representation, mainly to  local energy
fluctuations while the corresponding stochastic field acting on the
polaron leads mainly to  fluctuations of the renormalized transfer
integrals.

The exciton-phonon interaction in the ring subunits LH1 and LH2 of the
purple bacteria should not be very strong \cite{pull}.  Then the
Hamiltonian $H_{2} (t)$ models the influence of dynamic disorder mainly via
fluctuations of the local exciton energies $h_{mm}(t)$.  Mean values and
correlation functions are given by
\begin{eqnarray}
\langle h_{mm}(t) \rangle &=& 0,\nonumber\\
\langle h_{nn}(t)h_{mm}(\tau) \rangle
    &=&  \delta_{mn} \Delta^2 \exp(-\lambda |t-\tau|).
\label{corrf}
\end{eqnarray}
Multitime correlation functions entering the equations of motion
can be calculated from two-time correlation functions
(\ref{corrf})
according to different rules. Mostly dichotomic, Gaussian and
white noise statistics have been  applied in the past.

The simplest description of the exciton dynamics in the framework of the
stochastic theories is obtained in the so-called white noise limit
$\Delta^2/ \lambda \to  \gamma_0$ ($\Delta \to \infty$, $\lambda \to
\infty$) in which the time correlation functions become  $
\delta$ functions.  The SLE for the bare exciton density matrix in the
Haken-Strobl-Reineker parametrization (HSR-SLE) (for a review, see
\cite{reineker}) have been used  very often  to describe the coupled
coherent and incoherent regime of  exciton transfer in molecular
aggregates using a few phenomenological parameters $\gamma$:
\begin{eqnarray}
\frac{\partial}{\partial t}\rho_{mn}(t)&=&-\frac{i}{\hbar}\bigl(
[H_0,\rho(t)]\bigr)_{mn}\nonumber\\
&&+2 \delta_{mn}\sum_p[\gamma_{mp}\rho_{pp}
(t)-\gamma_{pm}\rho_{mm}(t)]\nonumber
\\
&&-(1-\delta_{mn})\bigl[2\Gamma_{mn}\rho_{mn}(t)
-2\bar\gamma_{mn}\rho_{nm}\bigr],\nonumber
\\
2\Gamma_{mn}&=&\sum_p(\gamma_{pm}+\gamma_{pn})=2\Gamma_{nm}.\label{SLE}
\end{eqnarray}
In most applications of the HSR-SLE nearest neighbor parameters $\gamma_{0}
= \gamma_{nm}$ for $n=m$, $\gamma_{1}=\gamma_{nm}$ for $n = m \pm{} 1 $ and
$\bar\gamma_{1}=\bar\gamma_{nm}$ for $n = m \pm{} 1 $ have been used.  We have
shown \cite{capbar} that in case of a weak interaction which is
site-diagonal and linear in the lattice displacements only $\gamma_{0} \neq
0$. Such treatment has been used by us \cite{barnato,barblumen,barcjp} and
Leegwater \cite{leeg} in the investigation of exciton transfer in ring
structures.

\subsection{Redfield theory}

Provided that the excitation dynamics is not very fast, the coupling is
rather weak, the bath and the system are initially statistically
independent, and except for the initial time-interval
$t\stackrel{<}{\sim}t_d$ ($t_d$~=~dephasing time of the bath), the
approximate equation turns to the form first derived by Redfield
\cite{redfield1,redfield2}:
\begin{equation}
 i\frac{d}{dt}\rho(t)=\frac{1}{\hbar}[H,\rho(t)]+{\cal R}\rho(t),
\label{eqmo}
\end{equation}
where the Redfield relaxation superoperator ${\cal R}$ (as a linear
time-independent superoperator) describes the influence of dynamics of the
medium and its interaction with the exciton.  In the modern fashion,
derivation of Eq.~(\ref{eqmo}) is perhaps best presented in \cite{blum96}
or \cite{may}. In the basis of eigenstates $|\mu{} \rangle,\,|\nu\rangle,\ldots$
of the relevant system part the Redfield theory reads
\begin{equation} \frac{d}{dt}\rho_{\mu{} \nu}(t)=
-\frac{i}{\hbar} E_{\mu{} \nu} \rho_{\mu{} \nu}(t) +\sum_{\mu{} '\nu'}
{\cal R_{\mu{} \nu\mu{} '\nu'}}\rho_{\mu{} '\nu'}(t) \label{eqredes} \end{equation}
where $E_{\mu{}  \nu}= E_\mu{} -E_{\nu} =\hbar\omega_{\mu{} \nu}$.
The physical interpretation of the so--called Redfield tensor ${\cal
  R_{\mu{} \nu \mu{} ' \nu'}}$, is straightforward: Each
element is essentially a rate constant that relates the rate of change of
reduced density matrix elements $\rho_{\mu{}  \nu}$ to the current value of the
other matrix element $\rho_{\mu{} ' \nu'}$.  Thus ${\cal R_{\mu{}  \mu{}  \mu{}  \mu{} }}$ is the
relaxation rate of the depopulation of eigenstate $|\mu{} \rangle$ while ${\cal
  R_{\mu{} \mu{} \nu \nu}}$ represents the rate of population transfer from state
$|\nu\rangle$ to the state $|\mu{} \rangle$.  The elements ${\cal
  R_{\mu{} \nu \mu{} \nu}}$ represent the dephasing rates of the off-diagonal
density matrix elements $\rho_{\mu{} \nu}$.  An important property of the
Redfield tensor is that when the bath is treated quantum mechanically, the
physically required detailed balance conditions are naturally satisfied, so
that the subsystem relaxes to thermal equilibrium with the bath. This
condition means that the ratio of forward and reverse rate constants
between states $|\mu{} \rangle$ and $|\nu\rangle$ is
\begin{equation}
  \frac{\cal R_{\mu{}  \mu{}  \nu \nu}}{\cal R_{\nu \nu \mu{}  \mu{} }} = \exp ( - \beta \hbar
  \omega_{\mu{} \nu})~. \label{detbal}
\end{equation}

Other elements of the Redfield tensor cannot be given such a simple
interpretation. Keeping them as just fitting semi-empirical parameters leads
to technical complications as the total number of elements of ${\cal R}$
increases as $N^4$ ($N$ is the number of eigenstates involved, i.e. two for
the dimer). Hence, the idea is to omit those elements of ${\cal R}$ which
are physically less relevant provided that this omission does not
appreciably influence the solution. That is why already Redfield himself
\cite{redfield1,redfield2} suggested the so--called secular approximation.
This approximation amounts to neglecting those elements of ${\cal R}$ that
couple elements of the density matrix $\rho_{\mu{} ' \nu'}$ with incommensurate
frequencies $E_{\mu{}  \nu}/\hbar$ determined in the first term on the right
hand side of (\ref{eqredes}). In other words, it means to
put $R_{\mu{}  \nu \mu{} ' \nu'}=0$ whenever $E_{\mu{}  \nu}\neq E_{\mu{} ' \nu'}$.  It has
been shown \cite{cbh,egorova} that this approximation leads to non-physical
effects in the transfer processes and is thus, in spite of its apparent
success in many cases, physically not acceptable if we  discuss,
e.g., relaxation among states other than eigenstates of the Hamiltonian of the
system.  Other approximations to the Redfield theory, which have not been
applied here, have been used and tested in the context of electron transfer
\cite{ukleine1,egorova}. Aspects concerning the efficient numerical
implementation of the Redfield equation are discussed elsewhere
\cite{ukleine2}.

Although being  aware that Redfield theory is limited to the high
temperature regime i.e.\ the system relaxation times have to be large
compared to the thermal time $\hbar /(k_b T)$ we assume here here that this
theory gives reasonable results also at low temperatures. In the present
study Redfield theory is applied to the full temperature range.

\subsection{Generalizations of the HSR-SLE method
in the local representation}

The original HSR-SLE method \cite{reineker} represents the simplest
stochastic treatment of the microscopic interaction Hamiltonian, but the
application of the white noise is very restrictive.  It leads to improper
long time asymptotics of the occupation probabilities of the extended
exciton states $P_{k}(\infty)$.  Many attempts have been made to determine
more general equations using different representations of the exciton
interacting with the bath and different ways of obtaining the dynamic
equations for the exciton density matrix from the full Hamiltonian
(\ref{he}).  The equations obtained in such a way could, after some further
approximations and manipulations, be rewritten \cite{cap1,cap3,cap4}
in a form which resembles the nonconvolutional SLE but brings several
important improvements.

\v{C}\'{a}pek \cite{cap1,cap3,cap4} applied several different ways of
obtaining of the convolutional and convolutionless dynamical equations for
the exciton density matrix.  The equations of motion obtained read
\begin{equation}
\frac{d}{dt}\rho_{mn}(t)=\sum_{pq}i\omega_{mn,pq}(t)\rho_{pq}(t),
\label{eqmot}
\end{equation}
where
\begin{eqnarray}
\omega_{mn,pq}(t)&=&\Omega_{mn,pq}+\delta\Omega_{mn,pq}(t)~.
\end{eqnarray}
When the typical system times are much longer than the bath correlation
time, i.e.\ in the Markovian limit, one obtain a system of equations with
constant coefficients
\begin{eqnarray}
  \Omega_{mn,pq}&=&\frac{1}{\hbar}[\delta_{mp}\delta_{nq}(\epsilon_n-
  \epsilon_m)+J_{qn}\delta_{mp}-J_{mp}\delta_{nq}],\nonumber\\
  i\delta\Omega_{mn,pq}&=&-\delta_{mp}{\cal
    A}_{nm}^q- \delta_{nq}{{\cal A}_{mn}^{p}}^{*},\nonumber
\end{eqnarray}
\begin{eqnarray}
{\cal
A}_{mn}^p&=&\frac{i\hbar}{N}\sum_k\omega_k^2(G_{-k}^m-G_{-k}^n)\sum_r
G_k^r\sum_{\nu_1,\nu_2}\langle\nu_2|r\rangle\langle
r|\nu_1\rangle\langle\nu_1|m\rangle\langle p|\nu_2\rangle\times{} \nonumber\\
&&\left\{\frac{1+n_B(\hbar\omega_k)}{E_{\nu_1}-E_{\nu_2}+\hbar\omega_k+
i\varepsilon}+\frac{n_B(\hbar\omega_k)}{E_{\nu_1}-E_{\nu_2}-\hbar\omega_k+
i\varepsilon}\right\}.
\end{eqnarray}
Here $|\nu\rangle$ and $E_{\nu}$ label the corresponding eigenvalues
and eigenvectors of $H_{ex}^0$.
\v{C}\'{a}pek suggested the parametrization
\begin{equation}
A_{mn}^p={\rm Re}\,{\cal A}_{mn}^p,\;B_{mn}^p={\rm Im}\,{\cal A}_{mn}^p.
\end{equation}
In the fast carrier regime the coefficients $B_{mn}^p$
become negligible (no renormalization of $J_{mn}$)
and can
be omitted.

The $A_{mn}^m$ play the same role as the coefficient
$\gamma_0\equiv\gamma_{mm}$ of the HSR parametrization. The coefficients
$A_{mn}^n$ provide the bath-induced coupling of to the off-diagonal to
diagonal elements of $\rho_{mn}(t)$.

\section{ Results}

\subsection{Equivalence of Redfield and \v{C}\'{a}pek's equations}

We have used the dynamical equations of motion for the
reduced exciton density matrix in the \v{C}\'{a}pek's form (25) and (26) to
express the time dependence of the optical properties of the model
LH2 ring of BChls in the fs time range.

For linear local exciton-phonon coupling
the Hamiltonian $H_{ex-ph}$ can be written as
\begin{equation}
H_{ex-ph}=\sum_m K_m\Phi_m,
\end{equation}
where  the
system operators $K_m$
and bath operators $\Phi_m$ are given by
$$
K_m = a_m^{\dagger}a_m, \qquad \Phi_m=
\frac{\hbar}{\sqrt{N}}\sum_{q}G_{q}^m\omega_{q}
(b^{\dagger}_{q}+b_{q}).$$

If we express the equation of motion for the reduced density matrix in Born
approximation (second order in exciton-bath coupling) in terms of
$K_m$ and $\Phi_m$, and apply the Markov approximation (in the interaction
picture), the dissipation term looks like \cite{may}
\begin{eqnarray} \label{dee}
\left (\frac{\partial \rho}{\partial t}\right
)_{diss}&=&-\frac{1}{\hbar^{2}} \sum_{m,n}\int_{0}^{\infty}d\tau\left
  \{C_{mn}(\tau)\left[K_m,K_n^{(I)}(-\tau)\rho(t)\right]_{-} \right .
\nonumber \\ &&~~~~~~~~~~~~~~~~~~ -\left .
  C_{mn}^{*}(\tau)\left[K_m,\rho(t)K_n^{(I)}(-\tau)\right]_{-}\right\},
\end{eqnarray} where $C_{mn}(t)$ is the bath correlation function
 \begin{equation}
C_{mn}(t)=\langle \Phi_m(t)\Phi_n(0)\rangle_{B}.
 \end{equation}
Here we assume that the expectation values $\langle \Phi_m \rangle$ vanish.
The original
Redfield equations are derived in eigenstate representation
\cite{redfield1,redfield2}. Here we will express Eq.~(\ref{dee}) in site
representation
\begin{eqnarray}
  \label{Redsr}
 \lefteqn{\left(\frac{\partial \rho_{rs}}{\partial t}\right )_{diss}=}
 \nonumber \\
&&-\frac{1}{\hbar^{2}}
\sum_{m,n}\int_{0}^{\infty}d\tau
\left\{
C_{mn}(\tau)
\sum_{i,j}\left(
\langle r|K_m|i\rangle\langle i|K_n^{(I)}(-\tau)|j\rangle\rho_{js}(t)-
\right. \right. \nonumber \\
&&\hspace{5.2cm} \left. \langle r|K_n^{(I)}(-\tau)|i\rangle\langle
j|K_m|s\rangle\rho_{ij}(t)\right)
 \nonumber \\
&&\hspace{2.8cm} -C_{mn}^{*}(\tau)
\sum_{i,j}\left(
\langle r|K_m|i\rangle\langle j|K_n^{(I)}(-\tau)|s\rangle\rho_{ij}(t)-
\right. \nonumber \\
&& \hspace{5.2cm}\left. \left.
\langle i|K_n^{(I)}(-\tau)|j\rangle\langle j|K_m|s\rangle\rho_{ri}(t)\right )
\right\},
\label{eqmsite}
\end{eqnarray}
where the indices $i$, $j$, $r$,
$s$ denote  sites.

For  $K_{m}$ as given above we have
\begin{equation}
 \langle r|K_m|s\rangle = \delta_{rm}\delta_{sm}~.
\label{km}
\end{equation}
Using the simple expression for the time
  evolution operator $U(\tau) $ in energy representation
  $\langle\alpha|U(\tau)|\beta\rangle=\delta_{\alpha\beta} {\mathrm
    e}^{-\frac{i}{\hbar}\varepsilon_{\alpha}\tau}  $
one gets
$$
\langle r|K_m^{(I)}(-\tau)|s\rangle =
\langle r|U(\tau)K_mU^{\dagger}(\tau)|s\rangle=
$$
\begin{equation}
\sum_{i,j}\sum_{\alpha,\beta}
\langle r|\alpha\rangle{\mathrm e}^{-\frac{i}{\hbar}\varepsilon_{\alpha}\tau}
\langle \alpha|i\rangle \delta_{in}\delta_{jn}
\langle j|\beta\rangle{\mathrm e}^{\frac{i}{\hbar}\varepsilon_{\beta}\tau}
\langle \beta|s\rangle=
\sum_{\alpha,\beta}
\langle r|\alpha\rangle\langle \alpha|n\rangle
\langle n|\beta\rangle\langle \beta|s\rangle
{\mathrm e}^{-i\omega_{\alpha\beta}\tau}.
\label{kimt}
\end{equation}
In this treatment the matrix elements of $K_m^{(I)}$ are evaluated exactly.
In the context of electron transfer the so-called {\it diabatic damping
  approximation} has been used and tested \cite{ukleine1,egorova}. In this
approximation the influence of the site coupling on dissipation is
neglected.

After substituting  Eqs.~(\ref{km}) and (\ref{kimt}) into (\ref{eqmsite})
the final form of the equation of motion for the reduced density
matrix in local site representation reads
\begin{eqnarray}
\label{localr}
\left (\frac{\partial \rho_{rs}}{\partial t}\right )_{diss} &=&
-\frac{1}{\hbar^{2}}\sum_{n,j}
\sum_{\alpha,\beta}\int_{0}^{\infty}d\tau
{\mathrm e}^{-i\omega_{\alpha\beta}\tau} \nonumber\\
&&\times{}
\left\{(C_{rn}(\tau)\right.-C_{sn}(\tau))
\langle r|\alpha\rangle\langle\alpha|n\rangle
\langle n|\beta\rangle\langle\beta|j\rangle\rho_{js}(t) \nonumber\\
&&-
(C_{rn}^{*}(\tau)-C_{sn}^{*}(\tau))\left .
\langle j|\alpha\rangle\langle\alpha|n\rangle
\langle n|\beta\rangle\langle\beta|s\rangle\rho_{rj}(t)
\right\}.
\end{eqnarray}

If one now introduces the  parameters ${\mathcal A}^m_{rs}$
\begin{equation} {\mathcal
  A}^m_{rs}=\frac{1}{\hbar^2} \sum_{\alpha,\beta}\int_{0}^{\infty}d\tau
{\mathrm e}^{-i\omega_{\alpha\beta}\tau}
\sum_n(C_{sn}^{*}(\tau)-C_{rn}^{*}(\tau)) \langle
m|\alpha\rangle\langle\alpha|n\rangle \langle
n|\beta\rangle\langle\beta|r\rangle \end{equation}
it is possible to express Eq.~(\ref{localr})
in the same form as it was done by
\v{C}\'{a}pek  \cite{cap4}:
\begin{equation}
  \left (\frac{\partial \rho_{rs}}{\partial t}\right
  )_{diss}=i\delta\Omega_{rs,pq} \rho_{pq}~,
\end{equation}
where
\begin{equation}
  i\delta\Omega_{rs,pq}=-\delta_{rp}{\mathcal A}^q_{sr}-\delta_{sq}
  {\mathcal A}^{p*}_{rs}~.
\end{equation}
So Redfield theory without secular approximation and \v{C}\'{a}pek's theory
after Markov approximation are equivalent.

In what follows we have used a simple model for $C_{mn}$. Each site
(chromophore) has its own bath with one bath coordinate, and the various
coordinates are completely uncorrelated.  Further it is assumed that these
baths have identical properties \cite{pull,may}. Then  only one
correlation function $C(\omega)$ is needed
\begin{equation} C_{mn}(\omega)=\delta_{mn}C(\omega),
\end{equation} with \begin{equation} C_{mn}(\omega)=\int_{-\infty}^{\infty}C_{mn}(\tau) {\mathrm
  e}^{i\omega\tau}d\tau= 2 {\rm Re} \int_{0}^{\infty}C_{mn}(\tau) {\mathrm
  e}^{i\omega\tau}d\tau.  \end{equation}

In case of real eigenvectors (which always can be chosen for this system)
$\langle m|\alpha\rangle$ and if
we omit the imaginary parts of ${\mathcal A}^m_{rs}$ which cause only
renormalization of transfer integrals $J$ \cite{cap4}, we can write
\begin{equation}
A^{m}_{rs}=Re{\mathcal A}^{m}_{rs}
=\frac{1}{2\hbar^2}\sum_{\alpha\beta}C(\omega_{\alpha\beta}) \langle
m|\alpha\rangle\langle\beta|r\rangle (\langle\alpha|s\rangle\langle
s|\beta\rangle- \langle\alpha|r\rangle\langle r|\beta\rangle) \end{equation}
 and after
introduction of the spectral density $J(\omega)$ \cite{may}
\begin{equation}
C(\omega)=2\pi\hbar^2[1+n_{B}(\omega)][J(\omega)-J(-\omega)], \end{equation} it is
possible to write
\begin{eqnarray}
  A^{m}_{rs}&=& -\pi\sum_{\alpha\beta}[1+n_{B}(\omega_{\alpha\beta})]
[J(\omega_{\alpha\beta})-J(\omega_{\beta\alpha})]  \times{} \nonumber \\
&& \langle m|\alpha\rangle\langle\beta|r\rangle
(\langle\alpha|s\rangle\langle s|\beta\rangle-
\langle\alpha|r\rangle\langle r|\beta\rangle) \nonumber \\
&=&-\pi\sum_{\alpha\beta(\varepsilon_\beta>\varepsilon_\alpha)}
J(\omega_{\alpha\beta})(\langle\alpha|s\rangle\langle s|\beta\rangle-
\langle\alpha|r\rangle\langle r|\beta\rangle)\times{} \nonumber\\
&&\left\{[1+n_{B}(\omega_{\alpha\beta})]
\langle m|\alpha\rangle\langle\beta|r\rangle+
n_{B}(\omega_{\alpha\beta})\langle m|\beta\rangle\langle\alpha|r\rangle
\right\}.
\end{eqnarray}

In the following we have modeled the spectral density by the form
usually used dealing with the BCHls in the LH2 ring unit
\cite{pull,kuehn97,may}
\begin{equation}
J(\omega)=\Theta(\omega)j_{0}\frac{\omega^{2}}{2 \omega_c^3}
{\mathrm e}^{-\omega/\omega_c}.
\label{jomega}
\end{equation}
It has its maximum at $2\omega_{c}$.  We shall use  (in
agreement with \cite{kuehn97}) $j_{0}= 0.2 $ or $j_{0}= 0.4 $ and
$\omega_c=0.2$.

\subsection{Disorder-induced scattering of the initial state}

In the unperturbed ring the time evolution of a exciton in the presence of
static site energy and dynamic disorder leads to a loss of its coherent
characteristics, i.e., destruction of the phase relationships between the
individual eigenstates comprising the initial wave packet state.  We have
calculated the time dependence of the occupation probability of the initial
state after impulsive excitation, namely
\begin{equation}
 P_{k}(t)= \rho_{kk}(t)
\end{equation}
for an equal superposition of $k= \pm{} 1$ states.  In the basis of the
unperturbed system the time evolution can be described as scattering of the
initially prepared exciton into the levels to which it is coupled by the
presence of disorder.  The scattering dynamics of a single ring after
averaging over more than 1000 realizations of the static disorder $\Delta$
is shown in Fig.~1 for four static disorders.  The influence of the dynamic
disorder is displayed for different  temperatures for three
values of dissipation.  The curves (i) with no dissipation do not coincide
with the corresponding results by Kumble and Hoch\-strass\-er, Fig.~2 in
\cite{hoch}. Their figures correspond to the time dependence of $P_{+1}(t)$
or $P_{-1}(t)$.  In Fig.~1 one can see a pronounced effect of the dynamic
disorder (curves (ii) - (v)) on the disorder-induced scattering of the
initial state, while the results without dynamic disorder (the Redfield
tensor set to zero) (i) coincide with those of Kumble and Hochstrasser
\cite{hoch} (when comparing the time dependence of $P_{+1}(t)$ or
$P_{-1}(t)$ with Fig.~2 in Ref.~\cite{hoch}).  With dissipation the
populations equilibrate while without dissipation the populations first
decay rapidly but then increase again. Without static and dynamic disorder
this would lead to full revivals. For short times some reminiscences of
these revivals are seen if one neglects dissipation. For low temperatures
 (cases (ii), (iv)), the dynamic disorder leads to a further decay
of the occupation probabilities $P_{1,-1}(t)$ after an initial fast decay.
Low temperature correspond to $k_B T/J_{12} = 0.001$ i.e.\ for $J_{12}=250
~{\rm cm}^{-1}$ the temperature is $T = 0.36K$.  For room temperature
the inclusion of dynamic disorder leads to faster initial decay.

\subsection{Consequences of the disorder-induced scattering
in optical anisotropy measurements}

The time-dependent anisotropy (\ref{aniso}) has been calculated for
excitation pulses polarized along the $e_x$ directions ($ \vec e_{i} = \vec
e_{x} $ ) and probe pulses which are polarized along the $e_x$ and $e_y$
directions ($ \vec e_{f} = \vec e_{x} $ and $ \vec e_{f} = \vec e_{y} $).
It is shown in Fig.~2.  Similar conclusions as for Fig.~1 can be drawn from
 Fig.~2. It was concluded in Ref.~\cite{hoch}
based on measurements by Chachisvilis et al.~\cite{chach}
that the
time decay of the anisotropy during
the first dozens of fs is
{ \it
  temperature independent} in the case of LH2 subunits.
Our calculation show that such
result  can  be obtained  only for $\Delta >0.8$.
But because
the time resolution of the experiments in \cite{chach} was not too high this
very restrictive statement about the strength of the static disorder
can only be made with caution. We expect that some temperature
dependence can be seen using experiments with shorter laser pulses.

\section{Discussion and concluding remarks}

We have used the LH2 ring structure of real antenna subunits of purple
bacteria photosynthetic systems as a motivation for our investigation of
the time dependent optical spectra of our molecular aggregate ring model.
In a series of papers the steady state optical absorption line shape of the
Frenkel excitons on cyclic model oligomers have been calculated
\cite{barmit,barchemphys}
in presence of both dynamic and quasistatic disorders. Correlation with
Freiberg's experimental data gives an estimate for the strength of the
static disorder $\Delta = 0.7$ \cite{barprehled}.

In the present investigation we study the time-dependent
optical properties under the influence of two simultaneous
site energy fluctuations - quasistatic and dynamic
within the LH2 ring antenna subunits of purple bacteria PSUs.
Despite a long history of experimental and theoretical investigations no
final conclusion has been done on the ratio $\Delta/J_{12}$.
Kumble and Hochstrasser concluded to reach a decay in time-dependent
optical experiments below 100 fs one needs $\Delta \approx 0.4 - 0.8
J_{12}$. They did not include dynamic disorder in their calculations. We
have extended those calculations by including dissipation. We were able to
show that in this case smaller values of  $\Delta/J_{12}$ are necessary
than predicted by  Kumble and Hochstrasser.

Our investigation has been based on several simplifications:

A) The time development of the reduced exciton density matrix has been
described by the long time limit of \v{C}\'{a}pek's dynamical equations of
motion (Markovian limit).  As shown in this paper  \v{C}\'{a}pek's equations
are equivalent to the
Redfield equation without secular approximation. More precise
description of the time development of the exciton density matrix during
first dozens of fs requires certainly more advanced description with time
dependent coefficients in Eq. (25) (non-Markovian treatment).
We had discussed such results
obtained by us only for a dimer using different methods for description of the
time development of the exciton density matrix \cite{bartai}.
In addition
we have assumed that Redfield theory could give  reasonable results for low
temperature.

B) We have used the ring symmetry to describe the polarization properties
of the Frenkel exciton states. Recent results obtained by single molecule
spectroscopy \cite{koehler} can only be interpreted \cite{knoemost}
admitting the presence of a $C_{2}$ distortion of the ring. It has,
up to now, not been concluded whether such $C_{2}$
distortion of the LH2 ring is present also in samples in vivo. Calculations
which take into account the possible $C_{2}$ distortion of the ring are on
the way.

C) Kumble and Hochstrasser have pointed out that the information needed for
a complete description of the exciton bath interaction is not available for
the LH2 units. We have worked with a form of the spectral density which has
been commonly used previously \cite{pull,kuehn97,may}.  Work on a better
description of the exciton--phonon interaction model is in progress.

\section*{Acknowledgements}

This work has been partially funded by contracts GA\v{C}R202/98/0499 and
GAUK345/1998 and by the BMBF and DFG. While preparing this work, I.B.
and P.H. experienced the kind hospitality of the Chemnitz University of
Technolgoy and U.K. experienced the kind hospitality of the Charles
University in Prague.

\newpage

{ \Large \bf Figure captions:}

\vspace{1cm}
\noindent
Fig. 1: Time dependence of the occupation probability $\langle P_{ \pm{}
  1}\rangle$ for static disorder $\Delta = 0.4$ (a), $\Delta = 0.6$ (b),
$\Delta = 0.8$ (c) and $\Delta = 1.0$ (d).  The influence of the dynamic
disorder is displayed by curves for low (ii), (iv) and room (iii), (v)
temperature for $j_{0} = 0.2$ (ii), (iii), and $j_{0} =0.4$ (iv), (v)
compared to $j_{0} = 0.0$ (i).

\psfig{figure=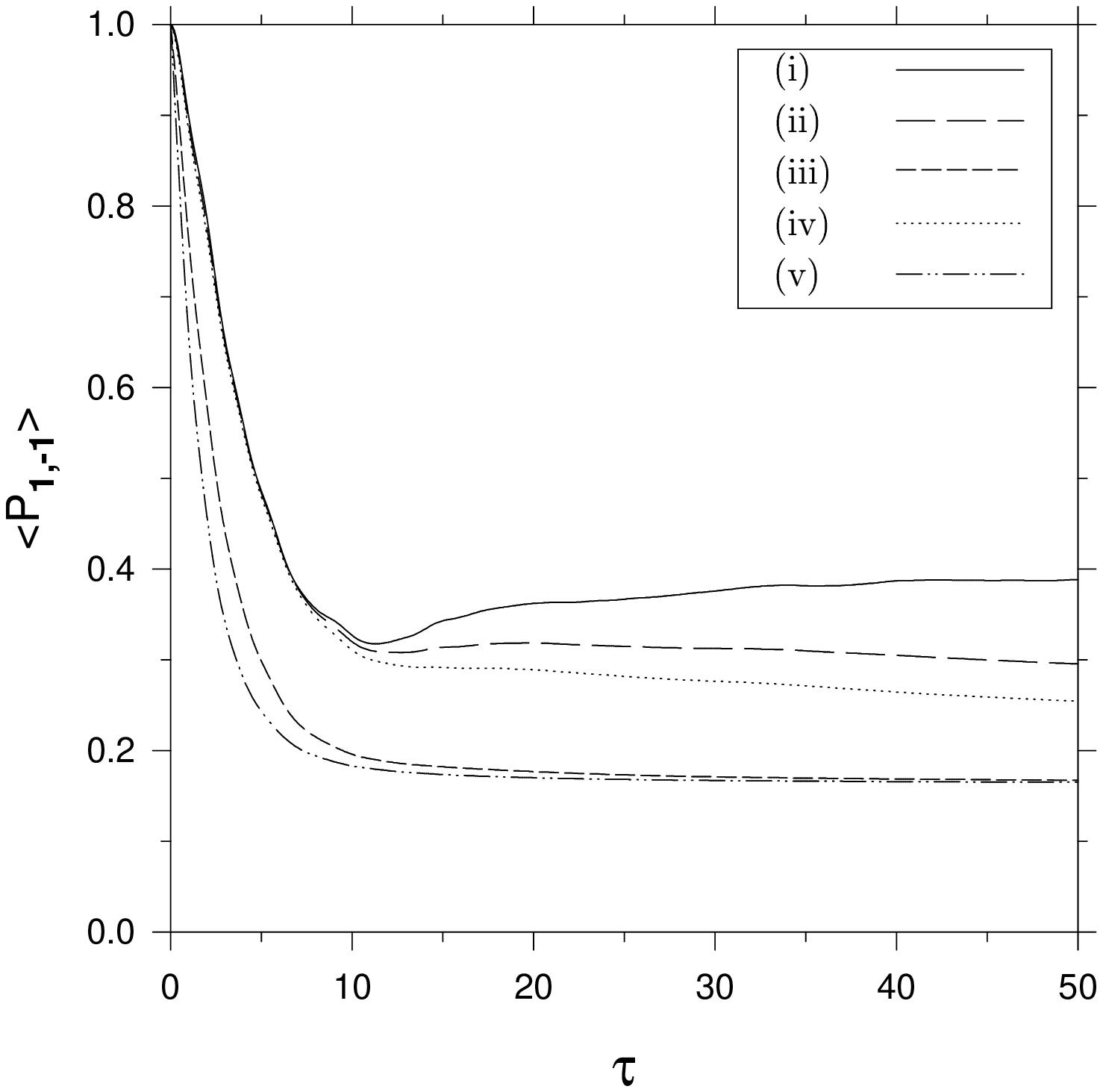,width=14cm}      
\psfig{figure=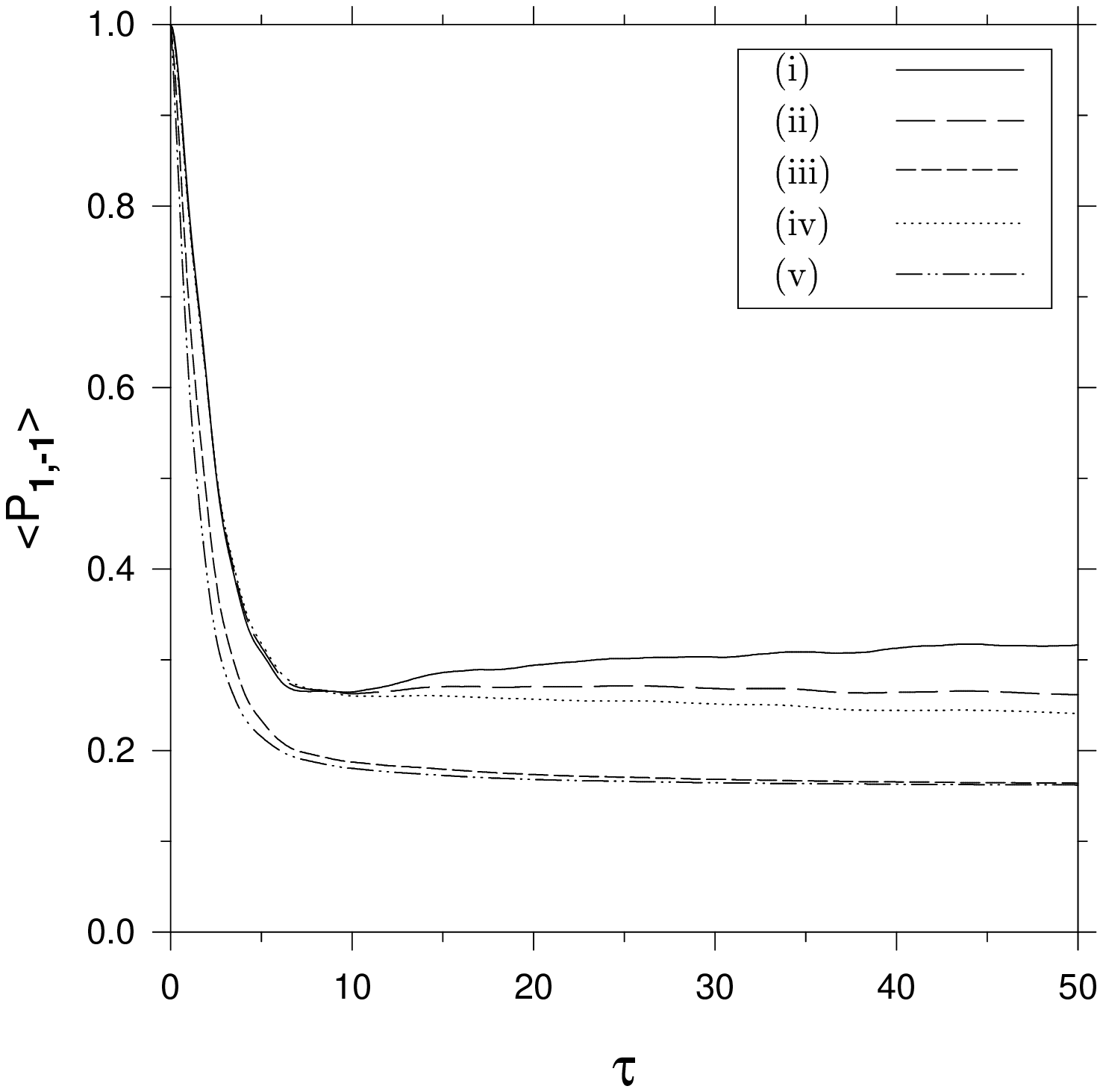,width=14cm}      
\psfig{figure=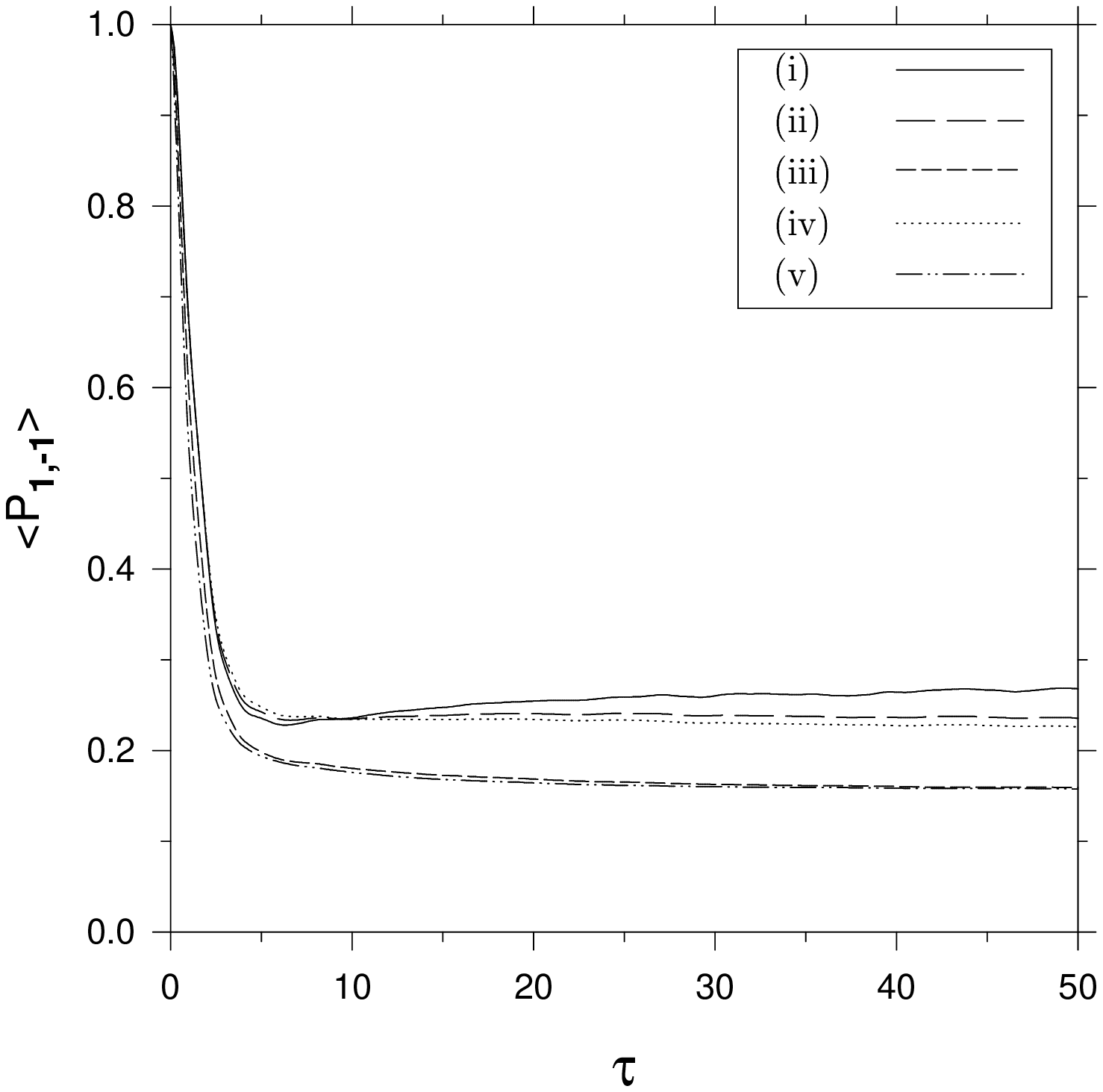,width=14cm}      
\psfig{figure=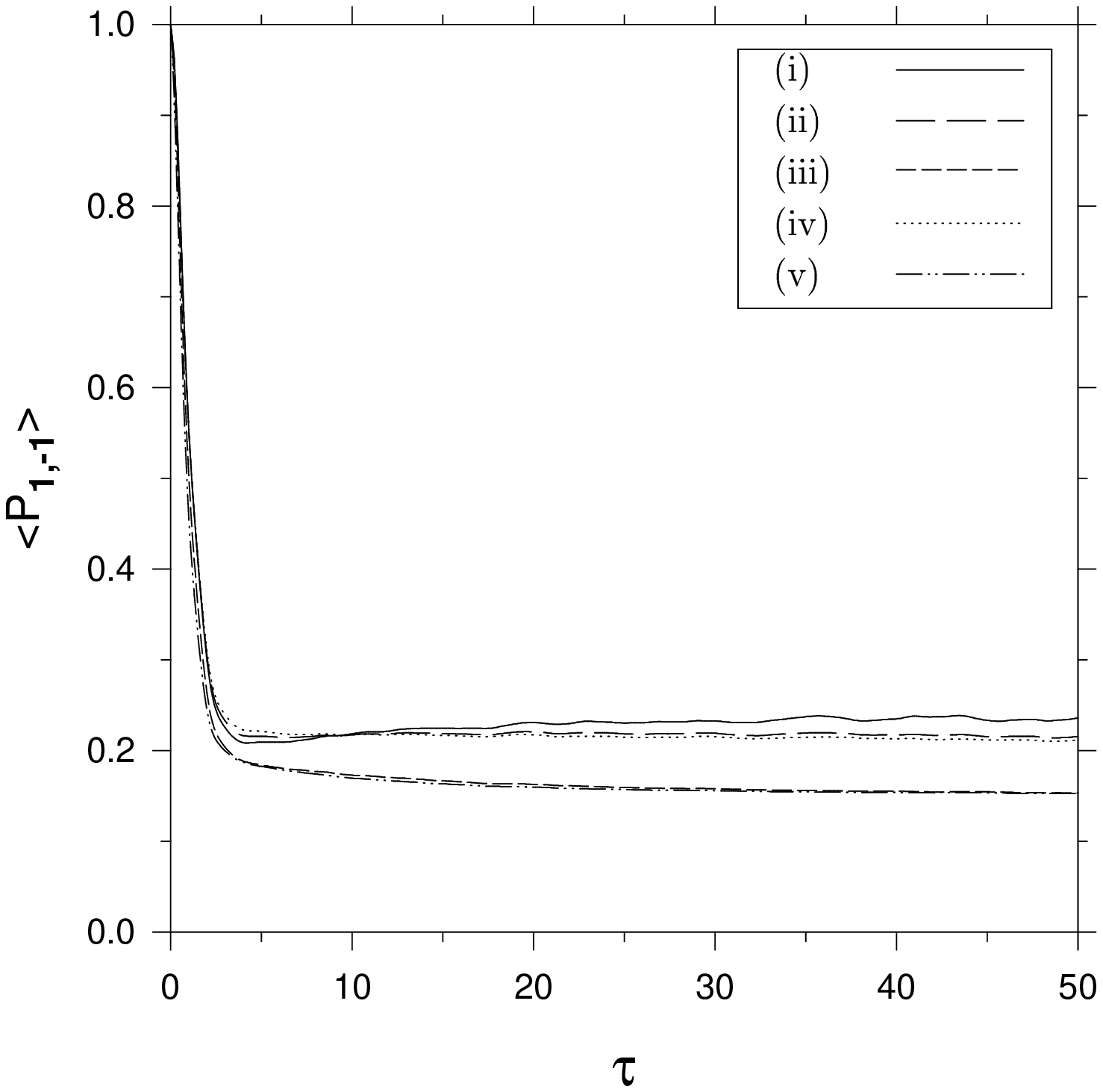,width=14cm}      

\vspace{1cm}
\newpage
\noindent
Fig. 2: Same as in Fig.~1 but for  the anisotropy $r(t)$.

\psfig{figure=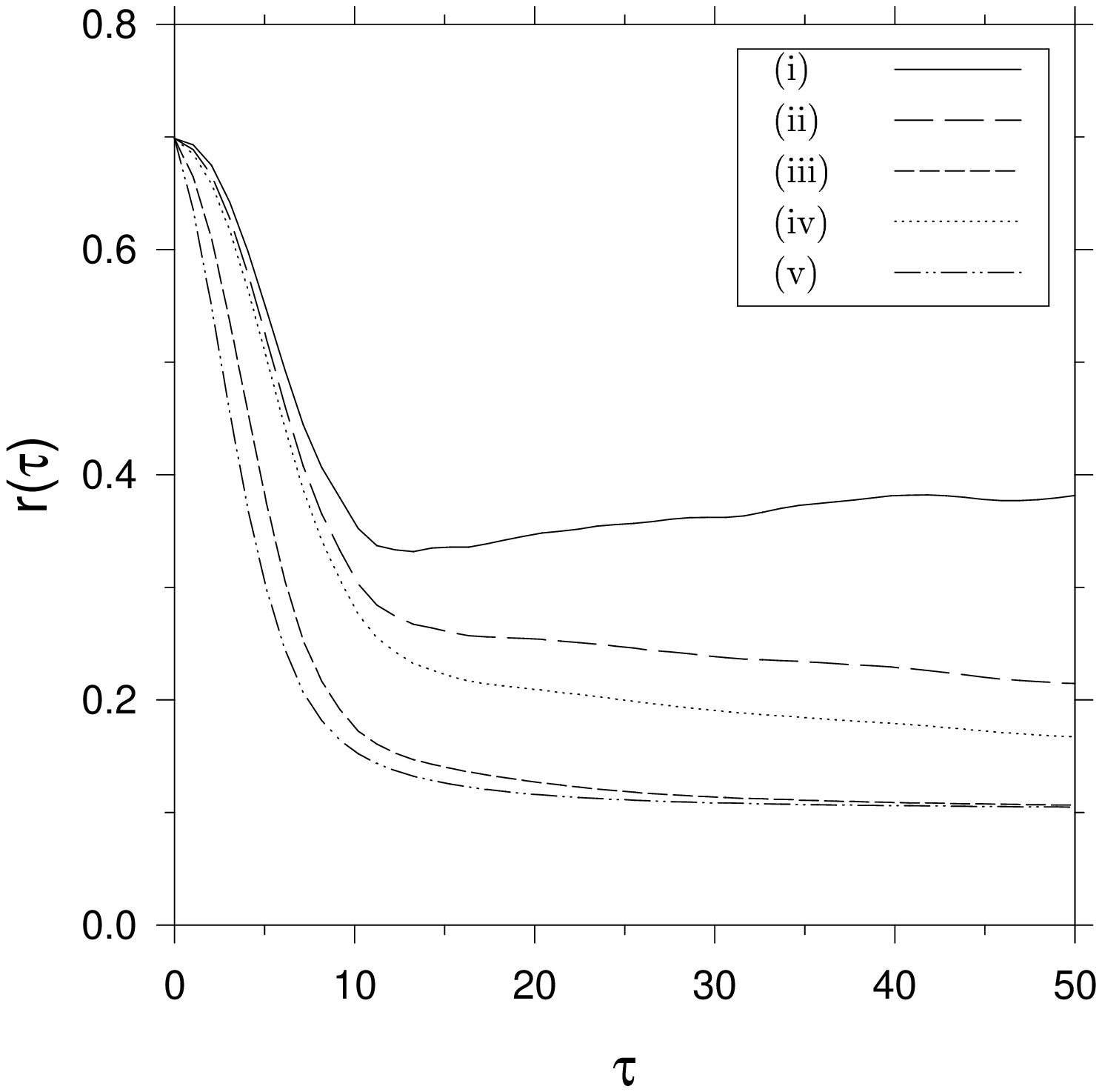,width=14cm}      
\psfig{figure=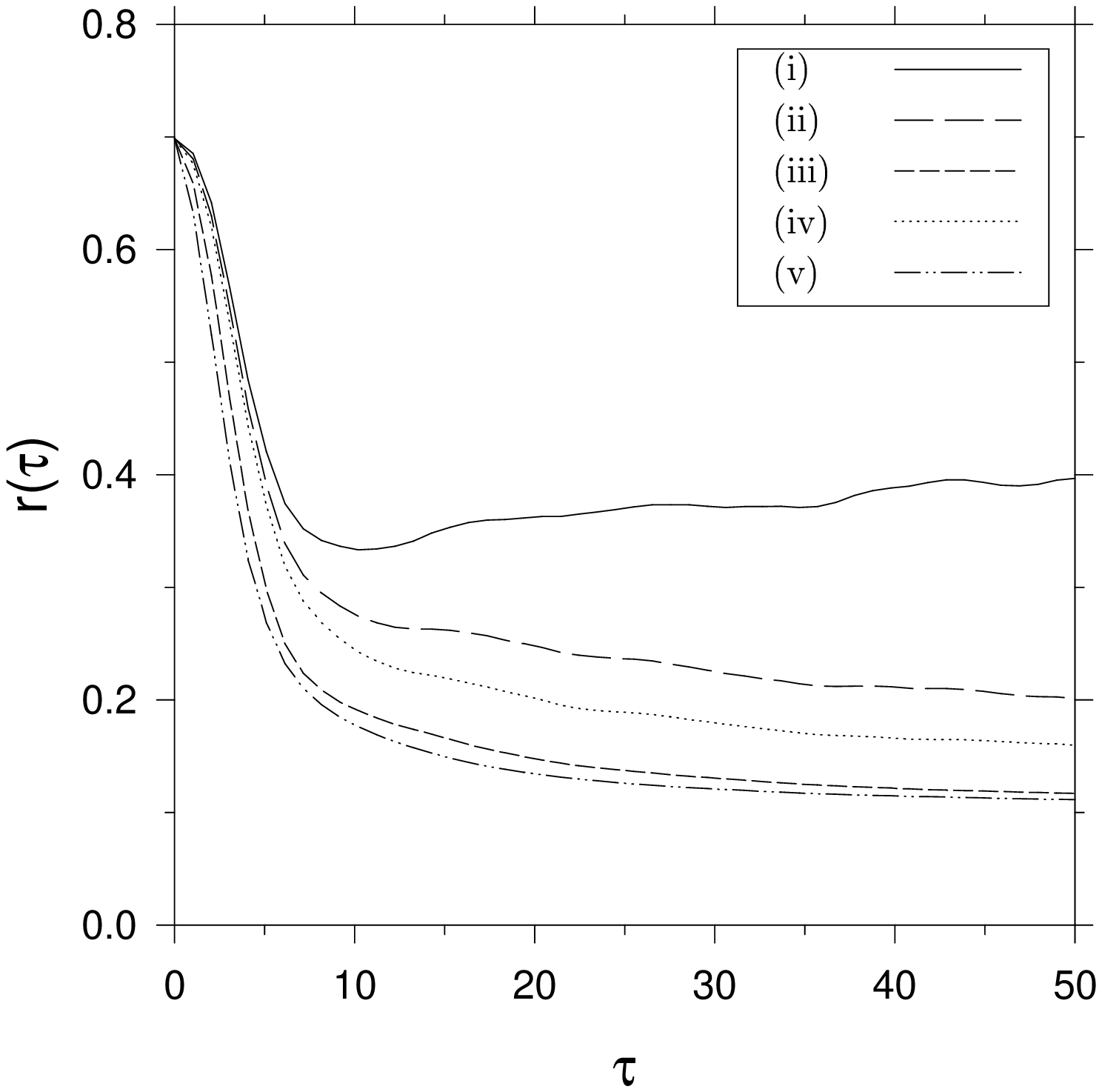,width=14cm}      
\psfig{figure=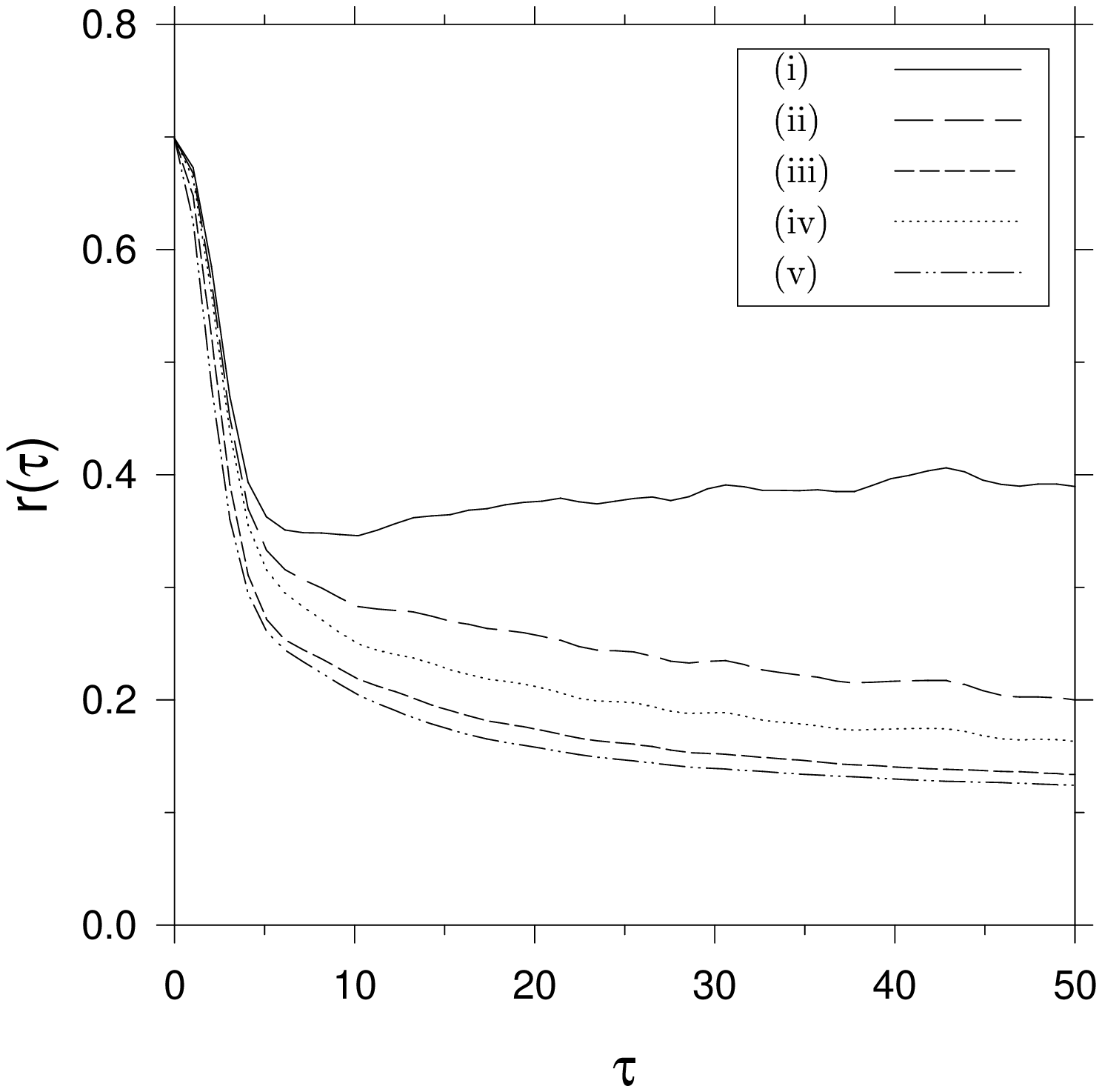,width=14cm}      
\psfig{figure=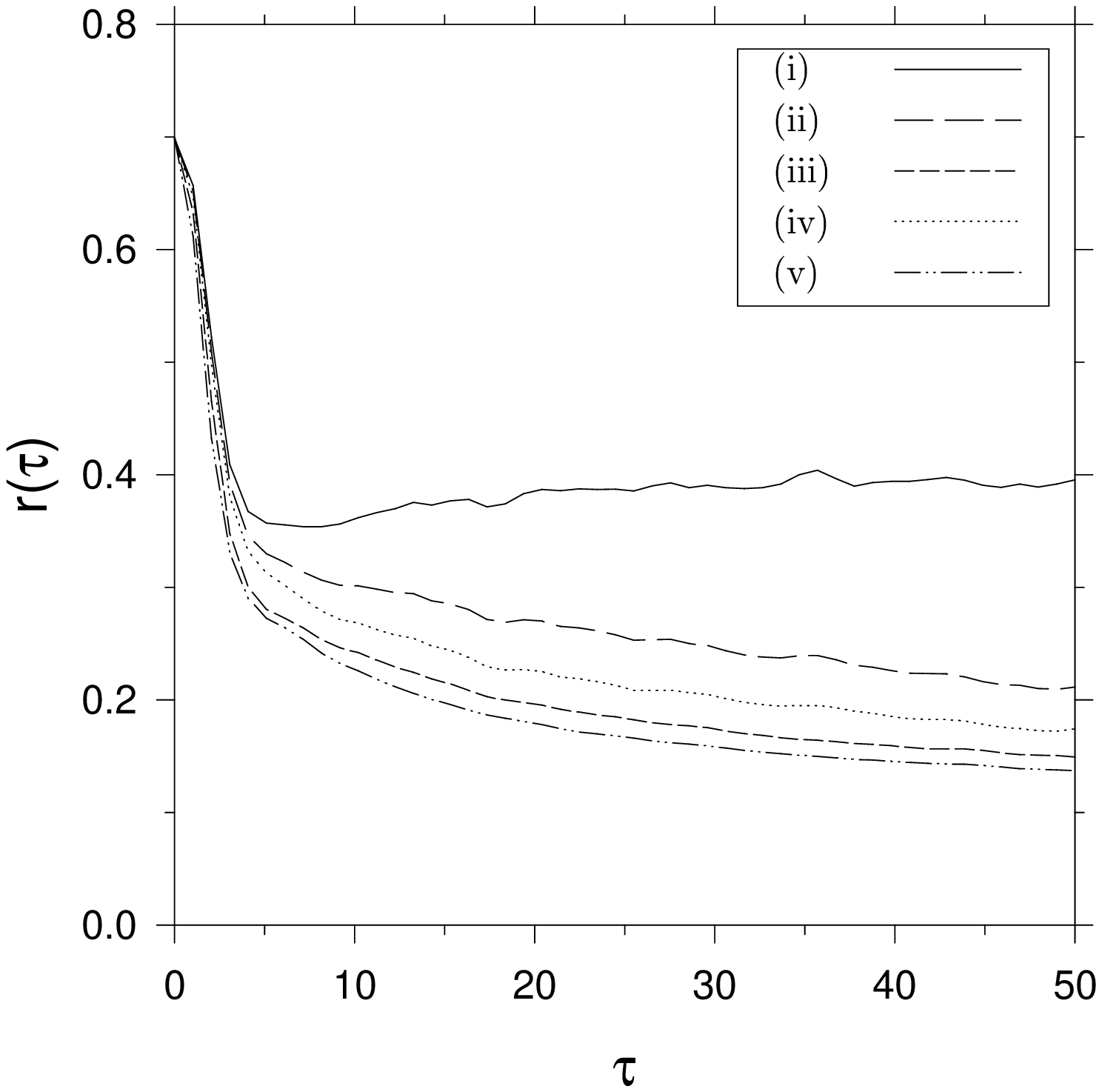,width=14cm}

\end{document}